\documentclass[onecolumn,preprint,amsmath,amssymb,epsf]{revtex4}
\usepackage{epsf}
\renewcommand{\vec}[1]{\mbox{\boldmath$#1$}}
\def\s{\sigma}              \def\g{\gamma}		
\def\D{\Delta}              \def\G{\Gamma}		
\def\k{{\bf k}}             \def\v{\vec{v}}		
\def\ls{\lambda_s}          \def\dmu{\delta\mu}		
\def\ls{\lambda_{s}}        		
\def\Dn{D^{\rm(n)}}   					
\def\timp{\tau_{\rm imp}}   
\def\timpN{\tau_{\rm imp}^{\rm(n)}}   
\def\tsf{\tau_{\rm sf}}

\def\tsfN{\tau_{\rm sf}^{\rm(n)}}
\def\RT{R_{\rm T}}          \def\RJ{R_{\rm J}}
\def\VT{V_{\rm T}}          \def\VJ{V_{\rm J}}
             
\def\S{{\cal S}}            \def\A{{\cal A}}           \def\/{\over}
\def\DN{{\cal D}_{\rm N}}   \def\DS{{\cal D}_{\rm S}}
\def\d{\partial}            \def\dE{\partial E}
         \def\dt{\partial t}

\begin{document}

\title{Spin transport and relaxation in superconductors}
\author{T. Yamashita,$^1$ S. Takahashi,$^1$ H. Imamura,$^2$ and S. Maekawa$^1$}
\address{$^1$Institute for Materials Research, Tohoku University, Sendai 980-8577, Japan} 
\address{${^2}$Graduate School of Information Sciences, Tohoku University, Sendai 980-8579, Japan}
\date{\today}


\begin{abstract}
We study theoretically the effect of spin relaxation on the spin
transport in a ferromagnet/superconductor(FM/SC) tunnel junction.
When spin-polarized electrons are injected into the SC from the FM,
nonequilibrium spin accumulation as well as spin current are
created in the range of the spin diffusion length in the SC.
We find that the spin diffusion length in the superconducting
state is the same as that in the normal state.
We examine a FM/SC/SC double tunnel junction, and show that 
the spin current is detected by the Joule heat generated at the
Josephson junction.  This provides a method to obtain the
spin diffusion length by probing the spin current in SC's.
\end{abstract}

\maketitle
\makeatother


Spin transport through magnetic nanostructures has attracted much interest.
Using the method of tunneling spectroscopy, Tedrow and Meservey \cite{meservey}
demonstrated that the current through the junction between a ferromagnet
(FM) and a superconductor (SC) is spin polarized.  Johnson and Silsbee
\cite{johnson,johnsonS} and Jedema {\it et al.} \cite{jemeda} have observed
nonequilibrium spin accumulation in a nonmagnetic metal sandwiched by
FM's.  Suppression of the superconducting gap due to spin accumulation has
been shown experimentally \cite{vasko,dong,yeh} and theoretically
\cite{takahashiPRL99}.

SC's are powerful probe for the spin polarization of the current
injected from FM's as shown in FM/SC tunnel junctions \cite{meservey}
and FM/SC point contacts \cite{soulen}.  
SC's are also useful for exploring
how the injected spin-polarized quasiparticles (QP's) are transported,
particularly the effect of spin relaxation on the spin transport,
because the spin-relaxation time and the spin diffusion length can be
measured precisely in the superconducting state where thermal noise
effects are extremely small.  
In addition, the unambiguous description of the spin-relaxation effect
is possible due to the fact that the spin relaxation is dominated
by spin-orbit impurity scattering in SC's.


So far, there have been a number of studies on the spin relaxation time and the
spin diffusion length in SC's.  However, the results are controversial: 
In a spin coupled resistance in permalloy/Nb/permalloy trilayers \cite{johnsonS}, 
it was shown that the spin diffusion length of Nb decreases with 
decreasing temperature in the superconducting state.  In contrast, 
the spin relaxation time in SC's was measured by the method of electron 
spin resonance (ESR) and was found to increase with decreasing 
temperature in SC's \cite{vier,nemes}. It was also shown that the spin 
diffusion length in SC's increases with decreasing temperature \cite{zhao1} 
by assuming that the length is proportional to the square root of 
the spin relaxation time.  Since the spin diffusion length and the 
spin relaxation time are key quantities for the spin transport in SC's, 
it is highly desired to construct a theory of the spin transport and 
relaxation and to solve the controversial issue mentioned above.

In this paper, we study the spin transport through a FM/SC tunnel junction.
The spin accumulation and spin current in SC are calculated based on
the Boltzmann transport theory.  It is shown that the spin diffusion
length in the superconducting state is equal to that in the normal state.
We examine a FM/SC1/SC2 double tunnel junction, and show that the spin
current is detected by the Joule heat generated at the Josephson
junction \cite{takahashiJAP01}, which provides information about
the spin diffusion length and the spin relaxation time in SC's.

We first consider a FM/SC tunnel junction.  The bias voltage $\VT$
is applied to the tunnel junction of resistance $R_{\rm T}$.
The tunnel barrier is at $x=0$ and the current flows in the $x$ direction.
The tunnel current is calculated by using the phenomenological tunnel
Hamiltonian which describes the transfer of electrons from one electrode
to the other.  If the SC is in the superconducting state, we rewrite
the electron operators $a_{{\bf k}\s}$ in the SC in terms of the
QP operators $\g_{{\bf k} \s}$ by using the Bogoliubov 
transformations, 
$a_{\k\uparrow} = u_{\bf k}\g_{\k\uparrow}
+ v_{\k}^*{\hat S}\g^\dagger_{-\k\downarrow}$ and $a^\dagger_{-\k\downarrow} 
= - v_{\k}{\hat S}^\dagger\g_{\k\uparrow}+ u^*_{\k}\g^\dagger_{-\k\downarrow}$, 
where $|u_{\k}|^2 = 1-|v_{\k}|^2={1\over 2}\left( 1+\xi_{\k}/E_{\k}\right)$,
${\hat S}$ is the operator which annihilates a Cooper pair \cite{josephson},
and $E_{\k}=[\xi_\k^2+\D^2]^{1/2}$ is the QP
dispersion with $\xi_{\k}$ and ${\D}$ being the one-electron
energy relative to the chemical potential of the condensate and the
superconducting gap, respectively.

From Fermi's golden rule, the spin-dependent tunnel currents
across the FM/SC junction are given by \cite{takahashiPRL99}
\begin{subequations}
\begin{mathletters}     
  \begin{eqnarray}      
    I_{\rm T\uparrow}(\VT) = \left({G_{\rm T\uparrow}/e}\right)
    \left[{\cal N} - \S(0) \right], 
    \label{eq:I1-u} \\
    I_{\rm T\downarrow}(\VT) = \left({G_{\rm T\downarrow}/e}\right)
    \left[{\cal N} + \S(0) \right], 
    \label{eq:I1-d}
  \end{eqnarray}        
\end{mathletters}       
\end{subequations}
where $G_{\rm T\s}$ is the tunnel conductance for electrons with spin
$\s$ when the SC is in the normal state, and $e$ the electronic charge.
The quantity ${\cal N}$ is the ordinary tunneling term driven by $\VT$:
${\cal N}(\VT) = \int_{-\infty}^\infty \DS(E)
  \bigl[ f_0(E-{e\VT}) - f_0(E)\bigr] dE$,
where $\DS(E)={\rm Re}\left[{|E|/\sqrt{E^2-{\D}^2}}\right]$ is the
normalized BCS density of states and $f_0(E)$ the Fermi distribution
function.  The quantity $\S(x)$ is the normalized spin density
at position $x$ in the SC;
\begin{equation}                              	
  \S= 1/(2{\cal D}_{\rm N}) \sum_{\k}			
  \bigl[f_{\k\uparrow}-f_{\k\downarrow}\bigr] ,	
  \label{eq:S}                               	
\end{equation}					
where $f_{\k\s} = \langle \g^\dagger_{\k\s} \g_{\k\s} \rangle$ is the
distribution function for a QP with energy $E_{\k}$ and spin
$\s$, and  $\DN$ the density of states in the normal state.
In Eq.~(1), we  neglected the contribution of charge imbalance by
assuming that the charge diffusion length $\lambda_Q$ is much smaller
than the spin diffusion length $\lambda_s$.  It has been reported that
$\lambda_s \sim 450~\mu$m (Ref. 2) and $\lambda_Q \sim 10~\mu$m
(Ref. 15) for Al.
The charge current   $I_{\rm charge}^{\rm T} = I_{\rm T\uparrow} + I_{\rm T\downarrow}$
and the spin current $I_{\rm spin}^{\rm T} = I_{\rm T\uparrow} - I_{\rm T\downarrow}$
through the junction are given by
\begin{subequations}
\begin{mathletters}
\begin{eqnarray}                      
  I_{\rm charge}^{\rm T}  = 
        \left[ {\cal N}- P\S(0) \right]/(eR_{\rm T}),\\
  I_{\rm spin}^{\rm T} = 
        \left[P{\cal N} - \S(0) \right]/(eR_{\rm T}),
  \label{eq:IqpIs}                   
\end{eqnarray}
\end{mathletters}
\end{subequations}
where $1/R_{\rm T} = G_{\rm T\uparrow} + G_{\rm T\downarrow}$ and
$P = (G_{\rm T\uparrow}-G_{\rm T\downarrow}) /
(G_{\rm T\uparrow}+G_{\rm T\downarrow})$ is the tunneling
spin polarization.

Let us examine the effect of spin relaxation on the spin accumulation 
and the spin current in SC.  In a steady state, the Boltzmann equation
is written as
\begin{eqnarray}
  {{\v_\k}}\cdot \nabla_{\bf r}
  f_{\k\s}+{\dot{\k}}\cdot \nabla_{\k}
  f_{\k\s}=\left( {\d f_{\k\s}}/{\dt}\right)_{\rm scatt},
  \label{boltzmann}
\end{eqnarray}
where $\v_\k=\hbar^{-1}\nabla_{\k}E_{\k}=(\xi_\k/E_\k)\v_{\rm F}$ is the
group velocity of QP's and $\v_{\rm F}$ the Fermi velocity.  
In the superconducting state, there is no electric field inside SC
and thus $\dot{\k}=0$.  The scattering term on the right side of
Eq.~(\ref{boltzmann}) arose from scattering of QP's
by nonmagnetic impurities, and is decomposed into the terms due to
elastic scattering and spin-flip scattering  \cite{bardeen}
\begin{eqnarray}
\left( {\partial f_{\k\s}}/{\partial t}\right)_{\rm scatt}
=-\frac{f_{\k\s}-f_{0\s}} {\timp}
 -\frac{f_{0\s}-f_{0-\s}} {2\tsf}  ,
\label{sfa}
\end{eqnarray}
where $f_{0\s}$ is the distribution function defined by the average
 of $f_{{\bf k}\sigma}$ with respect to the
direction of $\k$, $\timp=\left(E_\k/|\xi_\k|\right)\timpN$ \cite{bardeen}, 
and $\tsf=\left(E_\k/|\xi_\k|\right)\tsfN$ are the elastic and
the spin-flip scattering times in the superconducting state,
respectively, and $\timpN$ and $\tsfN$ are those in the normal state.

In the FM/SC junction, the physical quantities vary in the $x$
direction and are uniform in the $yz$ plane, where
$\nabla_y f_{\k\s}=\nabla_z f_{\k\s}=0$.  From
Eqs.~(\ref{boltzmann}) and (\ref{sfa}), we have
\begin{eqnarray}
  f_{\k\s} \sim f_{0\s}-{\timp}
  v_\k^x \nabla_x f_{0\s} .
  \label{nef}
\end{eqnarray}
The spin-dependent current density $i_{\s}$ flowing in the $x$
direction is calculated as
\begin{equation}
  i_{\s} = e\sum _{\k} v_\k^x f_{\k\s}
         = -2e\DN{\Dn} \int _{{\D}}^{\infty}
        \nabla_x f_{0\s} dE ,
  \label{SPC1}
\end{equation}
where $\Dn= \frac{1}{3} v_{\rm F}^2 {\timpN}$ is the diffusion constant
in the normal state. 

The spin accumulation at position $x$ in the SC is created by shifting
the chemical potential of up-spin QP's by $\delta\mu(x)$ and
that of down-spin ones by $-\delta\mu(x)$, which is described by taking
$f_{0\sigma}(E,x) = f_{0}(E-\sigma\delta\mu(x))$.
When $\dmu$ is much smaller than $\D$, $f_{0{\s}}$ is expanded as
\begin{eqnarray}
  f_{0\s}(E,x)  \sim f_{0}(E) 
  - {\s} \left[\d f_0(E)/\dE \right] \dmu(x) .
  \label{fe}
\end{eqnarray}
From Eqs.~(\ref{SPC1}) and (\ref{fe}),
the charge current density vanishes:
$i_{\rm charge}=i_{\uparrow}+i_{\downarrow}=0$, while the 
spin current density $i_{\rm spin}(x) =
  i_{\uparrow}-i_{\downarrow}$ is driven by the gradient of $ \dmu(x)$,
\begin{eqnarray}
  i_{\rm spin}(x) = -4e\DN \Dn f_0(\D)\nabla_x \dmu(x) ,
  \label{s1}
\end{eqnarray}
The divergence of $i_{\rm spin}(x)$ is given from
Eqs.~(\ref{boltzmann})-(\ref{fe}) by
\begin{equation}
  \nabla_x i_{\rm spin}(x)=
  -\left[{4e\DN}/{\tsfN}\right] f_0(\D) \dmu (x) .\label{s2}
\end{equation}
Thus, the chemical potential shift satisfies the equation
\begin{eqnarray}
    {\ls^2} \nabla_x^2 \dmu(x) = {\dmu(x)},
  \label{CPE}
\end{eqnarray}
where $\ls$ is the spin diffusion length in the SC
\begin{equation}
    \ls = \sqrt{ \Dn \tsfN }.
  \label{srln}
\end{equation}
In the FM/SC junction, Eq.~(\ref{CPE}) has a solution of the form
$\dmu(x) = \dmu(0) \exp(-x/\ls)$, and therefore both the spin
accumulation and spin current decay exponentially on the length
scale of $\ls$.  
{\em Note that the spin diffusion length in the superconducting state is
the same as that in the normal state.}  This result is understood as
follows: In the superconducting state, the diffusion constant is
$D = \frac{1}{3} v^2_\k {\timp} = \left(|\xi_\k|/E_\k\right) \Dn$
and the spin-flip time $\tsf=\left(E_\k/|\xi_\k|\right)\tsfN$,
so that the density of state factor $E_\k/|\xi_\k|$ in $\ls=\sqrt{D\tsf}$
is canceled out, resulting in Eq.~(\ref{srln}).

The spin injection experiment has been done to extract the spin diffusion
length $\ls$ in Nb by using bipolar spin transistors \cite{johnsonS}.
From the measurement of an excess voltage $V_s$ ($\propto \dmu$) due to
spin accumulation, a strong dependence of $V_s$ on temperature ($T$) was
found below the superconducting critical temperature $T_c$.
From an analysis of $V_s$ using the relation $V_s = V_{s0}{\rm exp}(-x/\tilde\ls)$,
$\tilde\ls \propto (1-T/T_c)^{-n}$ $(n \sim 1/2)$  was deduced \cite{johnsonS}.
However, since $\ls$ is independent of $T$ as given in Eq.~(\ref{srln}),
$\tilde\ls$ in Ref. 3 is not the spin diffusion length, but rather
the penetration length of the QP evanescent wave into SC due to Andreev
reflection (AR) \cite{blonder}.  This is because AR is dominant when SC is
in metallic contact with FM's as in the experiment of Ref. 3.
To measure $\ls$, it is desirable to insert a thin insulating barrier
between FM and SC for making the QP spin injection predominant.

\begin{figure}								
  \epsfxsize=0.9\columnwidth \centerline{\hbox{ \epsffile{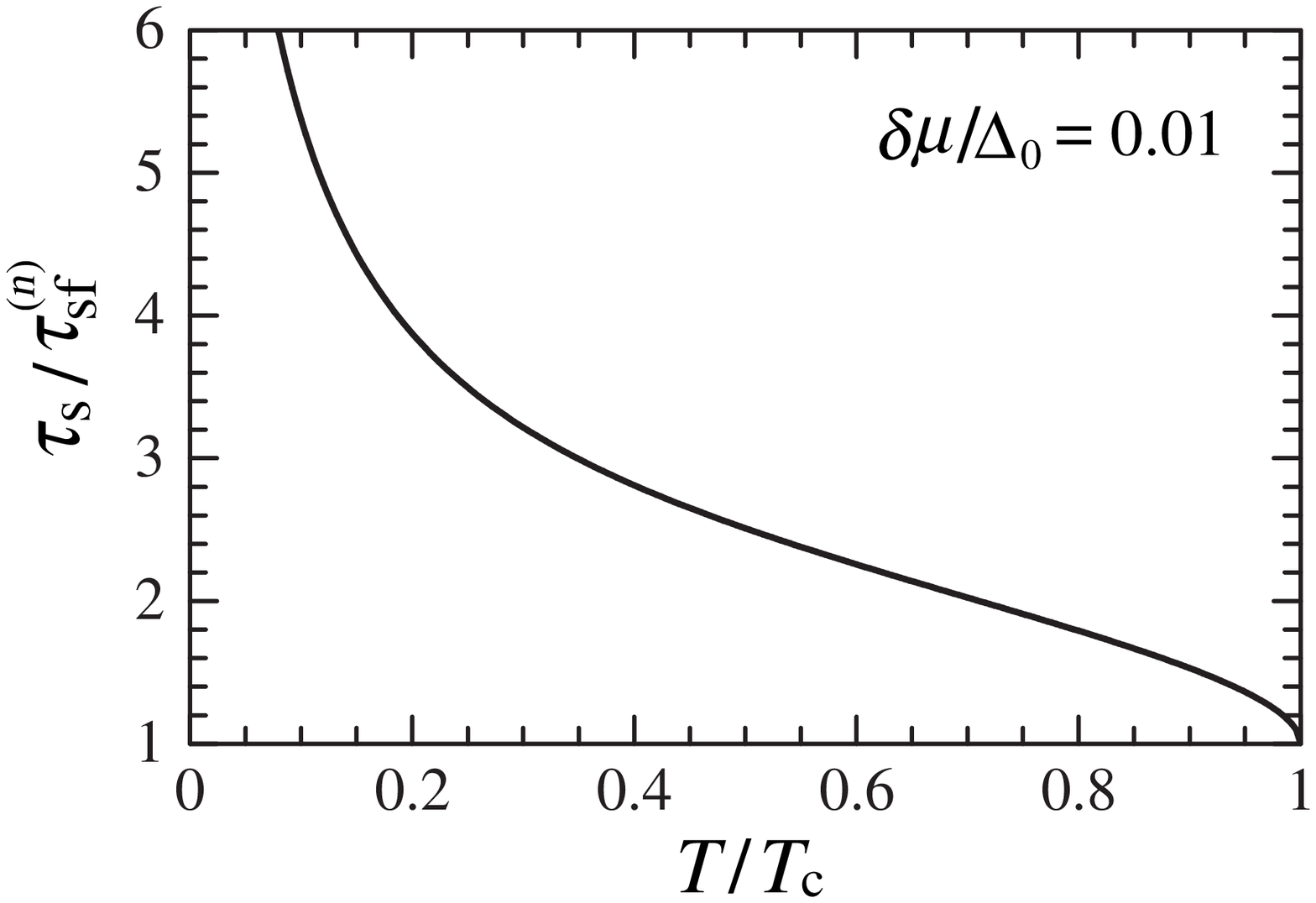}}}	
  \vskip 0.2cm
\caption{Temperature dependence of the spin relaxation time. 		
         The gap ${\D}_0$ is the value of $\D$ at $T=0$.}\label{fig1}	
\end{figure}

Another important quantity for the spin transport is the spin relaxation
time $\tau_s$ of $\S$ in the superconducting state, which is measured
by the ESR experiment.  If $\tau_s$ is introduced by the relaxation time
approximation $(\partial \S/\partial t)_{\rm scatt}=-\S/{\tau_s}$, we find
\begin{eqnarray}
  \tau_s=\frac{\displaystyle\int_{\D}^{\infty}
  \frac{|E|}{\sqrt{E^2-{\D}^2}}[f_{0\uparrow}-f_{0\downarrow}]dE}
  {\displaystyle\int_{\D}^{\infty}[f_{0\uparrow}-f_{0\downarrow}]dE}
  \tsfN .
 \label{taus}
\end{eqnarray}
For $\dmu \ll \D$, Eq.~(\ref{taus}) reduces to the result
of earlier theories \cite{yafet}.
Figure~\ref{fig1} shows the temperature dependence of $\tau_s$.
In the normal state ($\D=0)$ above $T_c$, $\tau_s$ is equal to $\tsfN$.
In the superconducting state below $T_c$, $\tau_s$ increases rapidly
with decreasing $T$ and behaves
similar to $\tau_s \simeq (\pi\D/2k_BT)^{1/2}\tsfN$ at low $T$.
It is worthwhile to note that one cannot use $\tau_s$ in place of
$\tsfN$ in Eq.~(\ref{srln}) when evaluating $\ls$, because $\tau_s$
is the relaxation time of the macroscopic quantity ${\cal S}$
while $\tsfN$ is the transport relaxation time of an individual
QP with particular energy, which makes them different
in the superconducting state.  In the normal state, however, $\tau_s$
is equal to $\tsfN$ and thus can be used for estimating $\ls$.
Note that $\delta\mu (0) \approx (\tau_{s}/\tau_{sf}^{(n)})
    \delta\mu^{(n)}(0)$ for $eV_T \ll \Delta$, where $\delta\mu^{(n)}(0)$
    is the shift of chemical potential in the normal state.

The above discussions are summarized as follows;
(1) The strong $T$ dependence of $V_s$ (Ref. 3) is not related
to $\ls$ but to the decay length of the evanescent wave in Andreev reflection.  (2) The ESR experiments \cite{vier,nemes} are consistent with our theory.  
(3) The theoretical treatment of $\ls$ in Ref. 12 
is not correct because they used the incorrect formula $\ls =\sqrt{\Dn\tau_s}$
which differs from Eq.~(\ref{srln}).

In order to investigate the spin diffusion length and the spin current
in SC's, we consider a FM/SC1/SC2 double tunnel junction.  The SC1 and SC2
are identical SC's, and their thicknesses are $d$ and semi-infinite,
respectively.
The resistance of the FM/SC1 tunnel junction and that of the SC1/SC2
Josephson junction (JJ) are $R_{\rm T}$ and $R_{\rm J}$, and the voltage
drops across the junctions are $\VT$ and $\VJ$, respectively.  The tunnel current
through the JJ is expressed as
\begin{eqnarray}                
 I =I_{\rm charge}^{\rm J}(\VJ) 
  +I_{\rm J1}(\VJ)\sin\varphi   
  +I_{\rm J2}(\VJ)\cos\varphi,  
    \label{eq:I}                
\end{eqnarray}                  
where $\varphi$ is the phase difference of the gap parameters in SC1
and SC2.  In Eq.~(\ref{eq:I}), the first term
describes the QP tunneling, and the second and third terms
describe the phase coherent (Cooper pair) tunneling.  The usual Josephson
effect is associated with the $\sin\varphi$ term.  Using Fermi's
golden rule, we have the spin-dependent QP tunnel current
\begin{eqnarray}                              
  I^{\s}_{\rm qp}(\VJ) = {1 \/ 2e R_{\rm J} }
    \int_{-\infty}^\infty dE \, \DS(E)\DS(E+e\VJ)
\left[ f_{0}(E-\s\dmu_1) - f_{0}(E+e\VJ-\s\dmu_2)  \right] ,
  \label{eq:Iqpus}                           
\end{eqnarray}
where $\dmu_i(x)$ is the shift of the chemical potential in the
$i$th SC.  The QP charge current
$I_{\rm charge}^{\rm J}=I_{\rm qp}^{\uparrow}(\VJ)
 + I_{\rm qp}^{\downarrow}(\VJ)$ and spin current
$I_{\rm spin}^{\rm J}= I_{\rm qp}^{\uparrow}(\VJ) 
 - I_{\rm qp}^{\downarrow}(\VJ)$ across the JJ are given by
\begin{eqnarray}
I_{\rm charge}^{\rm J}={1\/2e\RJ}
\int^{\infty}_{-\infty} dE \, {\DS(E)}{\DS(E+e\VJ)}
\sum_{\s=\pm} \left[ f_{0}(E-\s\dmu_1)-f_{0}(E+e\VJ-\s\dmu_2)\right] ,
\label{chcur}
\end{eqnarray}
\begin{eqnarray}
I_{\rm spin}^{\rm J}={1\/2e\RJ}
\int^{\infty}_{-\infty} dE \, {\DS(E)}{\DS(E+e\VJ)}
\Bigl[ f_0(E-\dmu_1)-f_0(E+\dmu_1) \nonumber\\
 - f_0(E+e\VJ-\dmu_2)+f_0(E+e\VJ+\dmu_2) \Bigr].
\label{spcur}
\end{eqnarray}
The phase coherent tunneling terms are obtained as
\begin{eqnarray}                                
  I_{\rm J1} = {\D^2\/2e\RJ}    
  \int_{-\infty}^\infty dE                      
  { \frac{\theta (|E|-\Delta)}{\sqrt{E^2 - \Delta^2}}\frac{\theta (\Delta - |E+e\VJ |)}{\sqrt{\Delta^2 - {(E+e\VJ)}^2}}}\nonumber\\
\times\sum_{j=1,2} \left[1 - f_{0}(|E|-\dmu_j)- f_{0}(|E|+\dmu_j)\right],      
  \label{eq:IJ1}
\end{eqnarray}
\begin{eqnarray}
  I_{\rm J2} = -{\D^2 \/2e \RJ } 
 \int_{-\infty}^\infty dE                       
  { \DS(E)\DS(E+e\VJ) \/ E(E+e\VJ)}
\sum_{\s=\pm} \left[f_0(E+e\VJ-\s\dmu_2)-f_{0}(E-\s\dmu_1) \right],
          \label{eq:IJ2}                                  
\end{eqnarray}                                  
where $\theta (x)$ is the step function.
Equations~(\ref{chcur})-(\ref{eq:IJ2}) are the generalized formulas
for the conventional JJ \cite{harris}. 
From Eq.~(\ref{CPE}), $\dmu_i$ has the forms 
$\dmu _{1}(x) = B_1 e^{x/{\lambda _s}} 
              + B_2 e^{-x/{\ls}}\label{CP1}$ in SC1
and $\dmu _{2}(x) = B_3 e^{-x/{\ls}}$ in SC2,
where $B_1$, $B_2$, and $B_3$ are determined by the boundary conditions
that the spin currents are continuous at each junction.
The results are used to calculate the currents through the FM/SC1/SC2 junction.

In the following we assume that the bias voltage across the JJ is zero
($\VJ=0$).  It follows from Eqs.~(\ref{chcur})-(\ref{eq:IJ2}) that
$I_{\rm charge}^{\rm J}$ and $I_{\rm J2}$ vanish, whereas
$I_{\rm spin}^{\rm J} \propto [\dmu_1(d) - \dmu_2(d)]$ and
$I_{\rm J1} \sim J_c$, $J_c$ being the Josephson critical current.
These results indicate that the charge current is carried by the Cooper
pairs as the dc Josephson current when the bias current is less
than $J_c$, while the spin current is carried by the QP's
as the QP current.

\begin{figure} [b]
  \epsfxsize=0.9\columnwidth \centerline{\hbox{ \epsffile{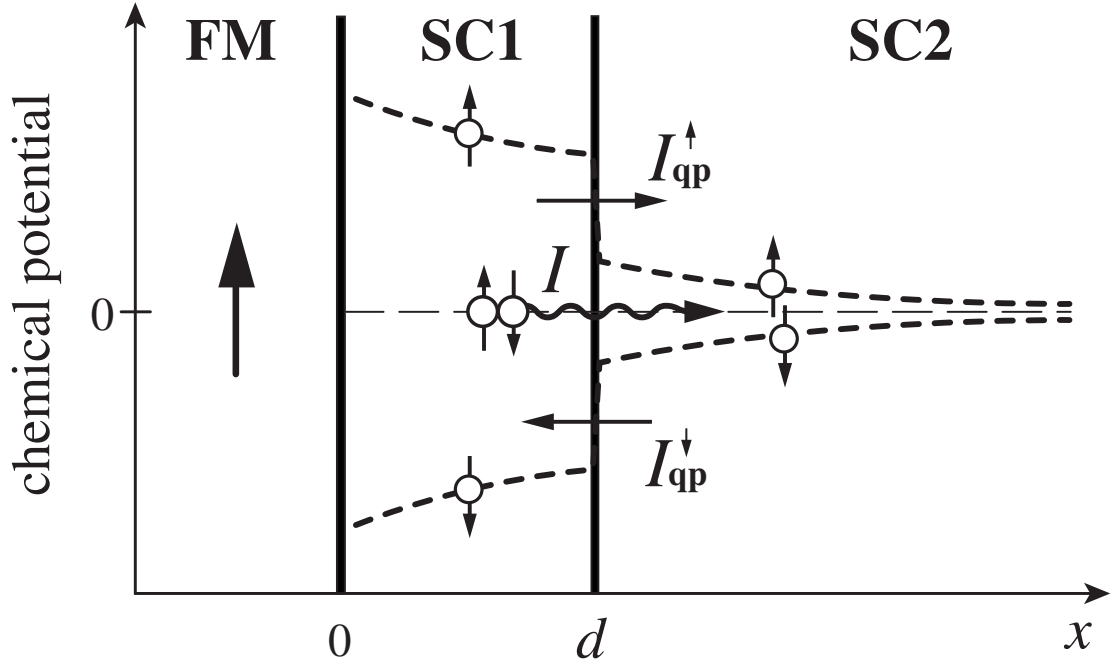}}}	
  \vskip 0.2cm
\caption{Spatial variation of the splitting in the chemical potentials	
of SC1 and SC2 in a FM/SC1/SC2 tunnel junction.  The dashed curves with	
up and down spins indicate the shifts, $\dmu_i(x)$ and $-\dmu_i(x)$, of
the up-spin and down-spin quasiparticles (QP's) in the $i$th SC, respectively,
and the long dashed line indicates the chemical potential of the Cooper 
pairs.}\label{fig2}							
\end{figure}

Figure~\ref{fig2} shows the spatial dependence of $\dmu_i$ ($-\dmu_i$)
for the up- (down-) spin QP's in the $i$th SC as well as the
pair and QP tunnel currents across the JJ at $\VJ=0$.
The up-spin tunnel current across the JJ is driven by the drop
$[\dmu_1(d) - \dmu_2(d)]$ in the forward direction,
while the down-spin one is driven by the same drop in the backward direction.
In the SC's, the up-spin and down-spin QP's, which are drifted
by the slope of the chemical potentials, flow in opposite directions
to each other, so that the QP's carry only the spin and do
not carry the charge.  This is one of the realizations of spin-charge
separation\cite{anderson}.

The most striking feature of the junction is that the spin current
across the JJ is accompanied by Joule heating at zero bias voltage
($\VJ=0$).   The power of Joule heating is given by
$W = \left[\dmu_1(d)-\dmu_2(d)\right] I^{{\rm J}}_{\rm spin}/e$,
and has the $d$ dependence of the form
\begin{eqnarray}                        
  W = {  P^2 \alpha \exp(-2d/\ls) / \left[1-\beta\exp(-2d/\ls)\right]^2} ,
\end{eqnarray}                          
where
\begin{eqnarray}                        
   \alpha = { 4\eta^2_1 [{\cal N}(\VT)/\G_0]^2 \G_2
    \/ e^2\RJ (1+\chi_1)^2(1+2\chi_2)^2 }, \ \
   \beta={(1-\chi_1)\/(1+\chi_1)(1+2\chi_2)} .
\nonumber
\end{eqnarray}                          
Here, $\chi_n = ({\G_n/\G_0})\eta_n$  ($n=1,2$), $\G_0 =  2f_0(\D)$, and
\begin{eqnarray}
  \G_n =  \int_{-\infty}^\infty
   \left[ \DS(E) \right]^n \left(-{\d{f_0}\/\dE}\right) dE ,
  \label{tau_n}
\end{eqnarray}  
with
 $\eta_1=(\rho_{\rm N}\ls / \RT\A)$,
 $\eta_2=(\rho_{\rm N}\ls / \RJ\A)$,
the normal-state resistivity $\rho_{\rm N}$ of SC,
and the junction area ${\A}$ \cite{hebel}.

\begin{figure}								
  \epsfxsize=0.9\columnwidth \centerline{\hbox{ \epsffile{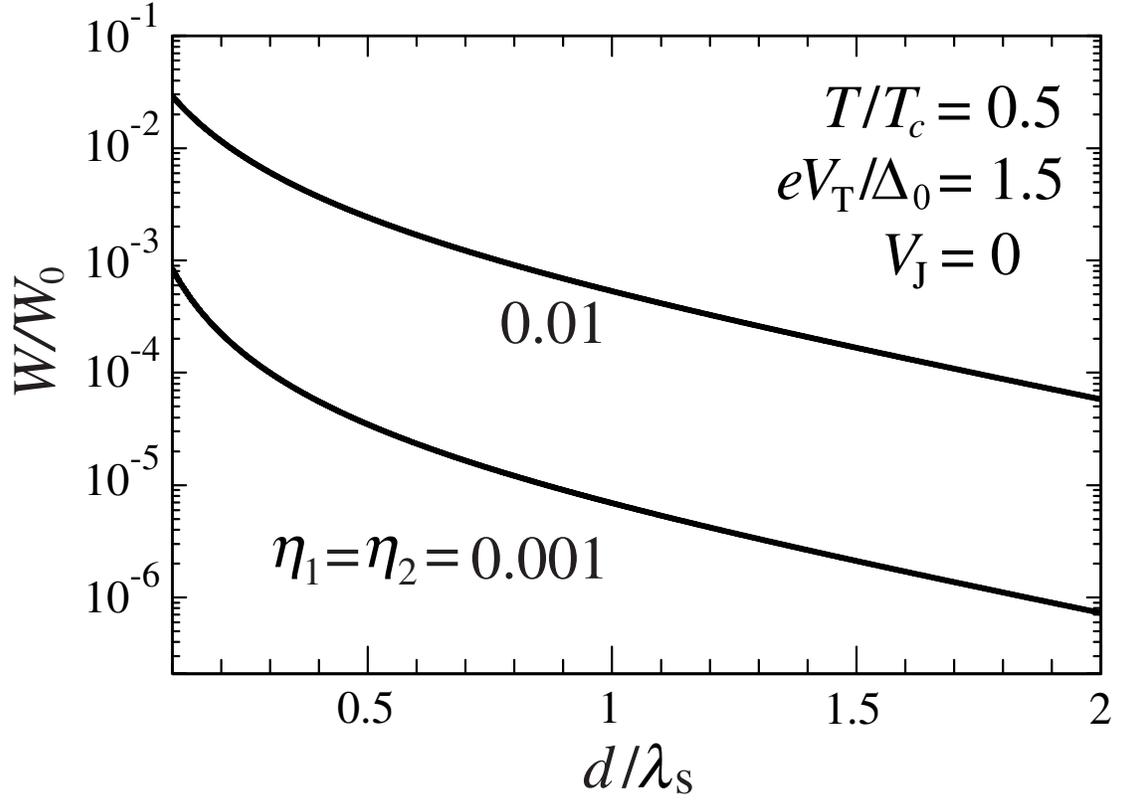}}}	
  \vskip 0.2cm
\caption{Joule heat $W$ generated at the JJ versus the thickness $d$	
of SC1.	 W is normalized by $W_0 = \D_0^2 /e^2 \RJ$.  }
 \label{fig3}		
\end{figure}

Figure~\ref{fig3} shows the Joule heat $W$ as a function of the
thickness $d$ of SC1 in the case where FM is a half metal ($P=1$).
An efficient generation of $W$ occurs for large values of $\eta_i$,
which corresponds to a low area tunnel resistance and/or long $\ls$.
It is seen that the curves show an exponential decay for
$d/\ls \agt 1$; $W \propto \exp(-2d/\ls)$.   At $d/{\ls} = 0.5$
and for $\eta_i = 0.01$, $\RJ\A = 10^{-6}~\Omega{\/}$cm$^2$, and
$\D_0=0.39$~meV (Al), we obtain $W/{\A} = 0.4$~mW/cm$^2$ per unit
area of the JJ, which is large enough to observe experimentally.
If $W$ is measured for various thickness of SC1 at $\VJ=0$, it
provides not only the spin diffusion length $\ls$ but also a
direct evidence for the spin current flowing in SC's.
Note that our method differs from the previous one \cite{johnson};
the former probes the {\it spin current} and the latter the
{\it spin accumulation}.

In conclusion, we have studied the effect of spin relaxation on the
spin transport in superconductors based on the Boltzmann equation,
and shown that the spin diffusion length in the superconducting
state is equal to that in the normal state.  This result resolves
the controversial issue of the spin diffusion length in the superconducting
state.  We propose a spin-injection device with the Josephson junction
to extract information about the spin diffusion length and the spin
current by measuring Joule heat generated at the Josephson junction.

This work was supported by a Grant-in-Aid from MEXT of Japan.
S.M. acknowledges the support of the Humboldt Foundation.

\newpage

\end{document}